\documentclass[11pt,a4paper]{article}
\usepackage{graphicx,amsfonts}

\newcommand{\be}{\begin{equation}}
\newcommand{\ee}{\end{equation}}
\newcommand{\bea}{\begin{eqnarray}}
\newcommand{\eea}{\end{eqnarray}}
\newcommand{\bdm}{\begin{displaymath}}
\newcommand{\edm}{\end{displaymath}}
\newcommand{\ul}{\underline}

\newcommand{\diag}{\mbox{diag}}

\newcommand{\diff}{d}
\newcommand{\Diff}{D}

\newcommand{\p}{\partial}

\newcommand{\Trace}{\mbox{Tr}}
\newcommand{\sprod}[2]{\langle #1\, , \,#2 \rangle}

\newcommand{\const}{\mbox{const.}}

\newcommand{\bbA}{\mbox{\boldmath $A$}}
\newcommand{\bbB}{\mbox{\boldmath $B$}}
\newcommand{\bbC}{\mbox{\boldmath $C$}}
\newcommand{\bbD}{\mbox{\boldmath $D$}}

\newcommand{\bbQ}{\mbox{\boldmath $Q$}}
\newcommand{\bbS}{\mbox{\boldmath $S$}}
\newcommand{\bbT}{\mbox{\boldmath $T$}}

\newcommand{\bbV}{\mbox{\boldmath $V$}}
\newcommand{\bbX}{\mbox{\boldmath $X$}}

\newcommand{\vep}{\ensuremath{\varepsilon}}

\newcommand{\taur}{\tau_r}

\newcommand{\gtens}{\mbox{\boldmath $g$}}

\newcommand{\gbartens}{\mbox{\boldmath $\bar{g}$}}

\newcommand{\ghat}{\hat{g}}
\newcommand{\gbar}{\bar{g}}
\newcommand{\Abar}{\bar{A}}

\newcommand{\nabhat}{\hat{\nabla}}
\newcommand{\asthat}{\hat{\ast}}

\newcommand{\lapbar}{\bar{\Delta}}
\newcommand{\nabbar}{\bar{\nabla}}

\newcommand{\Diffbar}{\bar{\Diff}}

\def\boxdal{\hbox{\hskip 0.5mm\hbox{\vrule width2.3mm height0.2mm
\vbox{\hrule width0.3mm height2.6mm}\hskip
-2.6mm \vbox{\hbox{\vrule width2.6mm height0.1mm}
\vskip -0.1mm\hrule width0.1mm height2.6mm}}\hskip 0.5mm}}

\begin{document}

\title{On the linear stability of solitons and
hairy black holes with a negative
cosmological constant: the odd-parity sector}

\author{O. Sarbach$^{\star}$ and E. Winstanley$^{\dagger}$\\
        $^{\star}$Center for Gravitational Physics and Geometry,\\
        Department of Physics, The Pennsylvania State University,\\
        104 Davey Laboratory, University Park, PA 16802.\\
        $^{\dagger}$Department of Applied Mathematics,
The University of Sheffield,\\
Hicks Building, Hounsfield Road, Sheffield, S3 7RH, U.K.}

\date{\today}

\maketitle

\begin{abstract}
Using a recently developed perturbation formalism based on
curvature quantities, we investigate the linear stability of black holes
and solitons with Yang-Mills hair and a negative cosmological
constant.
We show that those solutions which have no linear instabilities
under odd- and even-parity spherically symmetric perturbations
remain stable under odd-parity, linear, non-spherically symmetric
perturbations.
\end{abstract}

\section{Introduction}

The discovery of solitonic \cite{bartnik} and black hole
\cite{bizon} solutions to ${\mathfrak {su}}(2)$ Einstein-Yang-Mills
(EYM) theory sparked considerable study of the properties
of non-Abelian gauge theories coupled to gravity
(see \cite{review} for a comprehensive review of the subject and
a full list of references).
Many examples of both globally regular and black hole
solutions have now been discovered, in a wide range of
theories involving various non-Abelian
gauge fields, both with and without a (positive or
negative) cosmological constant.

Of this plethora of examples, many (but not all) are unstable.
In particular, there is a general result
that solitons and black holes solutions of EYM theory
with a compact gauge group in asymptotically
flat space must be topologically unstable \cite{brod2}.
This instability is analogous to the
instability of the flat-space Yang-Mills-Higgs
sphaleron \cite{galtsov}.
In addition, ${\mathfrak {su}}(2)$ EYM
solitons and black holes in the presence of
a positive cosmological constant are unstable \cite{brod1}.
It was therefore surprising to discover that
this result does not extend to solutions
in anti-de Sitter (adS) space, when
there is a negative cosmological constant.
Both soliton \cite{bjork} and black hole \cite{W-Stable}
solutions have been found which are linearly
stable with respect to spherically symmetric
perturbations.
It is the purpose of the present article to
examine the linear stability of these
solutions with respect to non-spherical
perturbations.

We concentrate here on the odd-parity
(or sphaleronic) sector.
For spherically symmetric perturbations,
the behaviour in this sector is ``topological''
in the sense that it does not depend on the
detailed structure of the solutions,
but only on global properties and boundary
conditions.
Further, it is in this sector (which,
for spherically symmetric perturbations, involves
the perturbations of the gauge field only, and
not the metric perturbations) in which the
analogy with the flat space sphaleron
is most pertinent.
Analysis of this sector for solitons and black holes
in asymptotically flat space has shown that
all the modes of instability are contained within
the spherically symmetric
perturbations \cite{S-Diss}.
Again, this is the same as for the flat space sphalerons,
where the only modes of instability are spherically
symmetric \cite{baacke}.
In this article we extend this result to solitons
and black holes in asymptotically anti-de Sitter space,
and show analytically that those solutions which are stable
under spherically symmetric perturbations
remain stable under non-spherically symmetric
perturbations in the odd-parity sector.
We shall return to the question of the stability
under non-spherically symmetric perturbations
in the even-parity sector in a future publication.

In order to discuss the stability, we
use a recently developed
perturbation formalism, which is based on curvature
quantities. The main advantage of this formalism is
that it allows us to cast the pulsation equations,
governing linear fluctuations on a static background,
into the form of a gauge invariant wave equation
even when complicated matter fields
are coupled to the metric.
It has been shown in \cite{BHS-Letter} that for a
static and purely magnetic solution of the EYM equations
(with arbitrary compact gauge group), the pulsation equations
admit the form of a symmetric wave equation for the
linearized extrinsic curvature and the electric field.
It is precisely this symmetric form of the pulsation
equations which will be crucial in this article to
show the stability of the above mentioned solitons and
black holes by analytic means.

This work is organized as follows.
In section \ref{Sect-2} we remember some important results about
the hairy black hole solutions with a negative cosmological
constant, recently found in \cite{W-Stable}, and
the corresponding solitonic solutions \cite{bjork}.
In section \ref{Sect-3}, we briefly review the curvature-based
formalism of perturbation theory for a static background,
and also generalize to the case where a cosmological constant is present.
The harmonic decomposition and the decoupling of the constraint
and dynamical variables are performed in section \ref{Sect-4},
where the absence of exponentially growing modes is also shown.
Technical details and the initial value formulation in terms of
gauge-invariant quantities are discussed in Appendix \ref{App-A}
and \ref{App-B}, respectively.

The metric signature is $(-,+,+,+)$ throughout, and we use the
standard notations $2\omega_{(ab)} = \omega_{ab} + \omega_{ba}$
and $2\omega_{[ab]} = \omega_{ab} - \omega_{ba}$ for
symmetrizing and antisymmetrizing, respectively.
Throughout the paper, greek letters denote
spacetime indices taking values
in $(0,1,2,3)$, while roman letters will denote spatial
indices taking values $(1,2,3)$.

\section{Solitons and
hairy black holes with a negative cosmological constant}
\label{Sect-2}

Hairy black holes in ${\mathfrak {su}}(2)$ Einstein-Yang-Mills
theory with a negative cosmological constant were first
found in \cite{W-Stable}, and subsequently in \cite{bjork},
where the corresponding solitonic solutions are also
discussed.
The equilibrium metric is spherically symmetric
\bdm
ds^{2}= -N(r)S^{2}(r)\, dt^{2}+N^{-1}(r)\, dr^{2} +r^{2}
\left( d\theta ^{2} +\sin ^{2} \theta \, d\phi ^{2} \right) ,
\edm
and the gauge field potential has the spherically symmetric
form
\bdm
A=(1-w(r)) \left[ -\tau _{\phi } \diff \theta + \tau _{\theta }
\sin \theta \, \diff\phi  \right] .
\edm
Here the ${\mathfrak {su}}(2)$ generators $\tau _{r,\theta ,\phi }$
are given in terms of the usual Pauli matrices $\sigma _{i}$
by $\tau _{r}={\ul {e}}_{r}\cdot {\ul {\sigma }}/2i$, etc.
Writing $N(r)=1-2m(r)/r-\Lambda r^{2}/3$, where $\Lambda $
is the (negative) cosmological constant,
the field equations take the form:
\bea
m_{r}& = & G\left[ Nw_{r}^{2} +\frac {1}{2r^{2}}(w^{2}-1)^{2}
\right] ,
\nonumber \\
\frac {S_{r}}{S} & = & \frac {2G w_{r}^{2}}{r} ,
\nonumber \\
0 & = & Nr^{2}w_{rr}+\left(
2m -\frac {2\Lambda r^{3}}{3} -
G \frac {(w^{2}-1)^{2}}{r} \right) w_{r} +(1-w^{2})w,
\label{eq-equil}
\eea
where $G$ is Newton's constant, and we have set the gauge
coupling constant equal to $\sqrt{4\pi}$ for convenience.

We are interested in solutions which approach anti-de Sitter (adS)
space at infinity, so that the asymptotic behaviour of the
field functions is:
\bea
m(r) & = & M + \frac {M_{1}}{r}+O(r^{-2}) ,
\nonumber \\
w(r) & = & w_{\infty }+\frac {w_{1}}{r} +O(r^{-2}) ,
\nonumber \\
S(r) & = & 1 + O(r^{-4}).
\nonumber
\eea
Here we already observe one difference between the
configurations in asymptotically flat, and asymptotically
adS space.
For asymptotically flat space solutions, either $w_{\infty}=0$,
in which case the Reissner-Nordstr\"om (RN) solution follows, or
$w_{\infty }=\pm 1$, so that there is no magnetic charge at
infinity \cite{ershov}.
However, in adS, the boundary conditions place no restriction
on the value of $w_{\infty }$, so in general, even non-Abelian
solutions will be globally magnetically charged.
For black hole solutions having a regular event horizon at
$r=r_{h}$, all the field variables have regular Taylor
expansions near the event horizon, for example
\bdm
w(r)=w(r_{h})+w_{r}(r_{h}) (r-r_{h})+O(r-r_{h})^{2} .
\edm
However, there are just two independent parameters, $S(r_{h})$
and $w(r_{h})$ since $N=0$ at the event horizon, which gives
\bdm
m(r_{h})= \frac {r_{h}}{2} -\frac {\Lambda r_{h}^{3}}{6}\, .
\edm
In order for the event horizon to be regular, we shall also require
that $N_r(r_h)>0$, which implies that
\bdm
F_h \equiv 1 - \Lambda r_h^2 - G\frac{(w(r_h) - 1)^2}{r_h^2} > 0.
\edm
From (\ref{eq-equil}), one has
\bdm
w_{r}(r_{h})=\frac {(w(r_{h})^{2}-1)w(r_{h})}{r_{h} F_h}\, .
\edm
There are also globally regular (solitonic) solutions,
for which the behaviour near the origin is:
\bea
m(r) & = & 2G b^{2} r^{3}+O(r^{4}) ,
\nonumber \\
w(r) & = & 1-br^{2}+O(r^{3}) ,
\nonumber \\
S(r) & = & S(0) \left[ 1+4G b^{2} r^{2} +O(r^{3}) \right].
\nonumber
\eea
Here the independent parameters are $b$ and $S(0)$.

The simplest solutions to the field equations (\ref{eq-equil})
are the Schwarzschild-adS solution,
\bdm
w=1, \;\;\; S=1, \;\;\; m=\const
\edm
and the RN-adS solution
\bdm
w=0, \;\;\; S=1, \;\;\; m=\const - \frac{G}{2r}\, .
\edm
In both cases, the YM field is effectively Abelian.
It is however interesting to study the stability of
these solutions with respect to non-Abelian perturbations.

The solutions of greatest interest in this article
are effectively non-Abelian solutions,
for which the gauge function $w$ has
no zeros, since these solutions were shown in
\cite{W-Stable} to be linearly stable to both
even and odd-parity spherical perturbations.
These solutions are unique to anti-de Sitter space,
as $w$ must have at least one zero if
the cosmological constant is positive or zero
\cite{breit,volkov}.
In \cite{W-Stable} it is proved that for any value
of the gauge field at the event horizon, $w(r_{h})\neq 0$,
for all sufficiently large $|\Lambda |$ there is a black hole
solution in which $w$ has no zeros.
Similar behaviour is found numerically for the solitonic
solutions \cite{bjork}.
For spherical perturbations of these equilibrium configurations,
it is proved analytically that all the solutions in which $w$ has
no zeros are stable in the odd-parity sector \cite{W-Stable}.
The even-parity sector is more complicated, but stability
can be proven for sufficiently large $|\Lambda |$.

In section \ref{Sect-4} we shall require various
properties of these soliton and black hole solutions,
particularly when $|\Lambda |\gg 1$.
We now briefly review the relevant results, and
refer the reader to \cite{bjork,W-Stable} for
further details and proofs.
We are concerned with black hole and soliton solutions
in which the gauge field function $w$ has no zeros,
which exist for sufficiently large $|\Lambda |$.
As $|\Lambda |\rightarrow \infty $, the function
$w$ approaches a constant value, given by $w(r_{h})$ for black
hole solutions, and $w\equiv 1$ for solitons.
This means that $w_{r}(r)$ tends to zero for all $r$ as
$|\Lambda |\rightarrow \infty $.
In \cite{W-Stable}, it was shown that in fact
\bdm
w_{r}(r) \sim o(|\Lambda |^{-\frac {1}{2}})
\qquad
{\mbox {as $|\Lambda |\rightarrow \infty $}}
\edm
for the black hole solutions, and that proof
is easily extended to show that the same
is also true for the solitons.
This result will be useful in section \ref{Sect-4}.

The other result we shall require in section \ref{Sect-4}
is that for these solutions in which $w$ has no zeros,
the gauge function $w$ cannot be equal to $\pm 1$ anywhere,
with the exception of the origin for the solitons
(where regularity conditions insist that $w=\pm 1$) and
the Schwarzschild-adS solution.
To see this, suppose that $w>0$ everywhere.
Then $w$ cannot have a local minimum in the region $0<w<1$
nor a local maximum in the region $w>1$ \cite{W-Stable,breit}.
For regular solutions, $w=1$ at the origin, and if it
is initially increasing, it is necessary for $w$ to first
reach a local maximum if it is to cross $w=1$ again, which
cannot happen.
Similarly, if $w$ is initially decreasing,
then, because $w$ has no zeros, it cannot have a local
minimum which is necessary if $w$ is to cross $w=1$ again.
For black hole solutions, the argument is dependent
upon the value of $w(r_{h})$.
If $w(r_{h})=1$, then $w\equiv 1$ for all $r$,
and the black hole geometry is Schwarzschild-adS.
If $w(r_{h})>1$, then initially $w$ is increasing, and the
same argument used above for the regular solutions applies.
If $w(r_{h})<1$, then initially $w$ is decreasing,
and once again $w$ cannot subsequently increase to cross $w=1$.
Exactly the same argument works in both cases if $w$ is
negative everywhere rather than positive.

In addition to the monopole solutions described in this
section, dyonic soliton and black hole solutions also
exist \cite{bjork}, again in contrast to the situation
in asymptotically flat space, when the electric charge
must be zero \cite{ershov}.
Due to the non-vanishing electric field, the stability
analysis of these solutions is more complex
(see, for example, \cite{kanti} for the stability
analysis of hairy black holes with non-vanishing
electric field), and in
this paper we shall focus only on the monopole solutions.
However, we would anticipate that those dyonic
solutions for which $w$ has no nodes would also be stable
in the odd-parity sector, at least when the electric
field is small enough.

\section{The pulsation equations}
\label{Sect-3}

We therefore consider linear fluctuations of a static and purely magnetic
soliton or hairy black hole solution in
Einstein-Yang-Mills  theory with a
cosmological constant.
In a recent letter \cite{BHS-Letter}, we have shown that,
at least in the pure EYM case, such fluctuations can be described by a
symmetric wave equation for the linearized extrinsic curvature and
the linearized electric field. In this section, we review the results obtained
there, and generalize to include a cosmological constant.
We discuss only the main steps in the derivation of a symmetric and
hyperbolic formulation for a static background and refer to
\cite{SHB-PRD} for more details.

One starts with the ADM equations of EYM$\Lambda$ theory, where the metric
and the gauge potential assume the form
\bea
\gtens &=& -\alpha^2 \diff t^2 + \gbar_{ij} (\diff x^i +
\beta^i \diff t)(\diff x^j + \beta^j \diff t),
\nonumber\\
     A &=& -\Phi\, \alpha\diff t + \Abar_i (\diff x^i + \beta^i \diff t).
\nonumber
\eea
($\Phi$ and $\Abar_i$ are both Lie algebra valued.)
On the background, the slicing $\Sigma_t$ is adapted to the staticity,
i.e. the shift $\beta$ and the time derivative of the 3-metric,
$\dot{\gbar}_{ij}$ vanish.
As a consequence, the extrinsic curvature tensor
\be
K_{ij} = \frac{1}{2\alpha} \left( \dot{\gbar}_{ij} - L_\beta\gbar_{ij} \right)
\label{Eq-ADMK}
\ee
vanishes on the background.
Similarly, the electric YM field, defined in terms of the field strength
$F = \diff A + A \wedge A$ and the future pointing normal unit vector field
orthogonal to the slices, $n = (\p_t - \beta)/\alpha$, by
\be
E_i = F_{i\mu} n^\mu,
\label{Eq-ADME}
\ee
is zero for static and purely magnetic solutions.
Since $K_{ij}$ and $E_i$ vanish on the background, the tensors
$\delta K_{ij}$ and $\delta E_i$, describing linear fluctuations,
are invariant with respect to both infinitesimal diffeomorphisms
within the slices and infinitesimal gauge transformation of the
gauge potential.
Hence, it is natural to look for a wave equation in terms of
these ``vector-invariant'' quantities.
In order to get such an equation, one then differentiates some evolution
equations with respect to the time coordinate $t$ and uses the
linearized version of equations (\ref{Eq-ADMK}) and (\ref{Eq-ADME})
in order to eliminate time derivatives of $\delta\gbar_{ij}$ and
the gauge potential $\delta\bar{A}_i\,$.
Next, one uses the linearized momentum and Gauss constraint equations
and spatial derivatives thereof in order to make the spatial operator both
elliptic and (formally) self-adjoint (the relevant scalar product
is given below).
Finally, one makes use of the freedom in choosing a (space-dependent)
reparametrization of the time coordinate in order to impose the maximal
slicing condition,
\bdm
\delta K^i_{\; i} = 0.
\edm
As a result, a hyperbolic and symmetric wave equation is obtained
from the following combinations of the field equations,
\bea
\Lambda_{ij} &\equiv& \frac{\alpha}{\sqrt{-g}}\,\p_t\,\delta\left(
\sqrt{-g}\, E_{ij} \right)
 - \frac{2}{\alpha} \nabbar_{(i} \left( \alpha^3 \delta E_{j)0} \right)
 + \frac{1}{\alpha^2} \gbar_{ij} \nabbar^k \left( \alpha^4 \delta E_{k0}
 \right), \nonumber\\
\Lambda^{(YM)}_i &\equiv& -\alpha\p_t \delta( \Diff^\mu F_{i\mu} )
                 + \frac{1}{\alpha}\Diffbar_i \alpha^3\delta(
\Diff^\mu F_{0\mu})
                 + 2\alpha^2\bar{F}^k_{\;\, i} \delta E_{k0}, \nonumber
\eea
where $E_{\mu\nu} = G_{\mu\nu} - 8\pi G T_{\mu\nu} = G_{\mu\nu}
- 8\pi G T^{(YM)}_{\mu\nu} + g_{\mu\nu}
\Lambda$.
Here and in the following, all quantities with a bar refer to the
background 3-metric $\gbar_{ij}$ and the
background magnetic potential $\Abar_i$\, .
Latin indices are raised and lowered with $\gbar_{ij}$, while
the index zero refers to the unit normal $n$.
Finally, $\Diff = \diff + [A |\, . \,]$ denotes the
covariant derivative with respect to the gauge potential.

The wave equation reads
\bea
0 &=& \hat{\Lambda}_{ij} = \hat{\Lambda}^{(vac)}_{ij}
+ \hat{\Lambda}^{(mat)}_{ij},
\nonumber \\
0 &=& \Lambda^{(YM)}_i,
\label{Eq-WaveEi}
\eea
where $\hat{\Lambda}_{ij}$ denotes the trace-less part  of
$\Lambda_{ij}$. In terms of the vector-invariant quantities
\bdm
L_{ij} \equiv \alpha\delta K_{ij}, \;\;\;
A \equiv \delta\dot{\alpha}, \;\;\;
{\cal E}_i \equiv \alpha\delta E_i\, ,
\edm
one finds, after using the background equations $E_{\mu\nu} = 0$
and $\Diff^\mu F_{\nu\mu} = 0$,
\bea
\hat{\Lambda}^{(vac)}_{ij} &=&  \boxdal_L L_{ij} + 4\nabbar_{(i}
\left( \alpha^k L_{j)k} \right)
    - 4\alpha_{(i} \nabbar^k L_{j)k} - 2\alpha\nabbar^k\!\left(
\frac{ \alpha_{(i} }{\alpha} \right) L_{j)k} \nonumber\\
 &  &- \frac{1}{\alpha} \nabbar_{(i} \alpha^2 \nabbar_{j)}\left(
\frac{A}{\alpha} \right)
    + \frac{1}{3} \gbar_{ij} \left( -\frac{2}{\alpha}
\nabbar^k(\alpha\alpha^l) L_{kl} + \lapbar A
    - R_{00} A \right),
\nonumber\\
\frac{ \hat{\Lambda}^{(mat)}_{ij} }{4G} &=&
\alpha\Trace\left\{ \bar{F}^k_{\;\, i} \bar{F}^l_{\; j} L_{kl}
 + \frac{1}{4} \bar{F}_{kl} \bar{F}^{kl} L_{ij} -
\frac{1}{3}\gbar_{ij} \bar{F}^{ks} \bar{F}^l_{\; s} L_{kl} \right\}
 - 8G\alpha \Lambda L_{ij} \nonumber\\
& &- \Trace\left\{ \Diffbar_k\left( \alpha\bar{F}_{(i}^{\;\; k} {\cal E}_{j)} \right)
 + \frac{ {\cal E}_k }{\alpha} \Diffbar_{(i} \left( \alpha^2\bar{F}_{j)}^{\; k} \right)
 + \frac{\alpha^2}{3}\gbar_{ij} \Diffbar_k\left(
\frac{ {\cal E}^l }{\alpha} \right) \bar{F}^{kl} \right\},
\nonumber\\
\Lambda^{(YM)}_i &=& \boxdal_E {\cal E}_i + 4G\alpha\Trace\left(
\bar{F}^{lj}{\cal E}_l \right)\bar{F}_{ij}\nonumber\\
 &  &+ 2\alpha\bar{F}^{jk}\nabbar_k L_{ij} -\frac{2}{\alpha} L_{kj}
\Diffbar^j(\alpha^2\bar{F}^{ki})
    - \alpha\bar{F}_{ij}\nabbar^j\!\left( \frac{A}{\alpha}\right),
\nonumber
\eea
where $\Trace$ stands for an Ad-invariant scalar product on the Lie
algebra and where all second partial derivatives are contained in
the hyperbolic operators $\boxdal_L$ and $\boxdal_E$, defined by
\bea
\boxdal_L L_{ij} &=& \left( \frac{1}{\alpha}\p_t^{\; 2} -
\nabbar^k\alpha\nabbar_k \right) L_{ij}
    + 2\alpha\bar{R}^k_{\; (i} L_{j)k} - 2\alpha\bar{R}_{kilj} L^{kl},
\nonumber\\
\boxdal_E {\cal E}_i &=&  \frac{1}{\alpha}\ddot{{\cal E}}_i
+ 2\Diffbar^j\left( \alpha\Diffbar_{[i} {\cal E}_{j]}\right)
    - \frac{1}{\alpha}\Diffbar_i\alpha^3\Diffbar^j\!\left(
\frac{ {\cal E}_j}{\alpha} \right)
    -\alpha [ \bar{F}_{ij}, {\cal E}^j]. \nonumber
\eea
Note that the spatial parts of the operators defined in
(\ref{Eq-WaveEi})
are symmetric with respect to the scalar product
\bea
&& \sprod{ (L^{(1)}, {\cal E}^{(1)}) }{ (L^{(2)}, {\cal E}^{(2)})} \nonumber\\
&& \quad\quad\quad\equiv\int\limits_\Sigma \left[
  \gbar^{ik}\gbar^{jl} L^{(1)}_{ij} L^{(2)}_{kl}
+ 2G\Trace\left\{ \gbar^{ij} {\cal E}^{(1)}_i  {\cal E}^{(2)}_j \right\}
\right] \sqrt{\gbar}\,\diff^3 x.
\label{Eq-InnerProduct}
\eea
The constraint equations are the linearized momentum constraint,
\be
0 = \alpha\delta E_{i0} =
\alpha\nabbar^j\!\left( \frac{L_{ij}}{\alpha} \right)
- 2G\Trace\left( \bar{F}^j_{\; i}{\cal E}_j \right),
\label{Eq-LADMEYMLMomentumCons}
\ee
and the linearized Gauss constraint
\bdm
0 = -\alpha\delta( \Diff^\mu F_{0\mu} ) =
\alpha\Diffbar^j\!\left( \frac{ {\cal E}_j }{\alpha} \right).
\edm
Additional constraints involving also perturbations of the metric and the
gauge potential themselves are the Hamiltonian constraint and
all evolution equations,
which we had differentiated in time in order to construct the
wave operator.

Since we adopt the maximal slicing condition, there is an elliptic
equation for $A$, which is obtained from the trace of the tensor
$\Lambda_{ij}\,$,
\be
\left( \lapbar - R_{00} \right)A =
   2\nabbar^k\left\{ \alpha^l L_{kl} + G\Trace(
\alpha\bar{F}_{lk}{\cal E}^l) \right\}
   - 2G\alpha\Trace \left\{ \bar{F}^{km}\bar{F}^l_{\; m} L_{kl} \right\},
\label{Eq-LADMEYMLElliptic}
\ee
where the momentum constraint equation (\ref{Eq-LADMEYMLMomentumCons})
and the background equations have been used in order to simplify
the equation.
Here $R_{00}$ is the $00$-component of the Ricci tensor.
From the background equations, one finds that
$R_{00} = G\Trace( \bar{F}^{kl}\bar{F}_{kl} )/2 - \Lambda$,
and therefore, for a negative $\Lambda$, the operator on
the left-hand side of equation (\ref{Eq-LADMEYMLElliptic})
is negative and the equation is solvable.
As we have argued in \cite{SHB-PRD}, this equation decouples
when one projects the wave operator onto the momentum constraint manifold.
In the case of odd-parity perturbations on a spherical symmetric background,
as we are going to study below, equation
(\ref{Eq-LADMEYMLElliptic}) is in fact void
since it is a scalar equation.

\section{Odd-parity fluctuations}
\label{Sect-4}

\subsection{The form of the perturbation equations}

We now apply the curvature-based perturbation theory
in order to study the
linear stability of the solutions discussed in section
\ref{Sect-2}.
Since the background is spherically symmetric in this case,
it is convenient to expand the linearized extrinsic curvature
and electric YM field in terms of spherical tensor harmonics.
Perturbations belonging to different
choices of the angular momentum numbers $\ell$ and $m$ decouple.
Furthermore, the tensor harmonics can be divided into parities.
Here, we consider only odd-parity (axial) perturbations.
The even-parity (polar) case will be subject to a future article.
Since the perturbations of all scalar quantities vanish for
odd-parity perturbations, the variation of the lapse and the trace of
the extrinsic curvature do not appear and therefore the
amplitudes parametrizing $\delta K_{ij}$ and $\delta E_i$ are
fully coordinate- and gauge-invariant.

How to expand $L_{ij}$ and ${\cal E}_i$ in spherical tensor
harmonics and how to find the corresponding expansion of the wave
operator is explained in Appendix~\ref{App-A}. We introduce a
coordinate $\rho $ such that $\p_\rho = NS \p_r\,$.
Since we are dealing with geometries which are asymptotically adS,
$\rho$ does not tend to infinity as $r\rightarrow \infty $, but
instead $\rho $ tends to a finite value $\rho _{max}$. We shall
use the notation $\rho _{0}$ to denote the lowest value of $\rho
$, which is $0$ at the origin for solitonic solutions, and
$-\infty $ at the event horizon for black holes. Using this
``tortoise'' coordinate, the resulting wave equation is
\be
\left( \p_t^{\, 2} - \p_\rho^{\, 2} + \bbA\p_\rho + \p_\rho\bbA
+ \bbS \right) U = 0,
\label{Eq-SymWaveEYMLOdd}
\ee
where $U \equiv (h,a,b,k,c,d,e)$,
with $h$ and $k$ parametrizing metric
perturbations and $a$, $b$, $c$, $d$ and $e$ parametrizing the YM
field.
The matrices $\bbA$ and $\bbS$ are antisymmetric and
symmetric, respectively, and have the block structure
\bdm
\bbA = \left( \begin{array}{cc} \bbA_1 & 0 \\
0 & \bbA_2 \end{array} \right), \;\;\;
\bbS = \left( \begin{array}{cc} \bbS_1 & \bbS_t^T \\
\bbS_t & \bbS_2 \end{array} \right),
\edm
where $\bbA_1$ and $\bbS_1$ are $3\times 3$ matrices, whereas
$\bbA_2$ and $\bbS_2$ are $4\times 4$ matrices.
Defining
\bdm
\gamma \equiv \frac{\alpha}{r}, \;\;\;
f \equiv w^2 + 1, \;\;\;
u \equiv \frac{w_\rho}{r}, \;\;\;
v \equiv 2\sqrt{G}\,\frac{w^2-1}{r}\, ,
\edm
the non-vanishing matrix elements of $\bbA_1$ and $\bbA_2$
can be written as
\bea
&& (\bbA_1)_{13} = -(\bbA_1)_{31} =-\sqrt{G}\, u,
\nonumber\\
&& (\bbA_2)_{14} = -(\bbA_2)_{41} =-\sqrt{2G}\, u.
\nonumber
\eea
The matrix $\bbS$ is given by
\bea
\bbS_1 &=& \gamma^2\left( \begin{array}{ccc}
 \frac{r}{\gamma^3} \left( \frac{\gamma}{r} \right)_{\rho\rho}
+ \lambda + v^2 & sym. & sym. \\
  \mu\, v & \frac{\gamma_{\rho\rho}}{\gamma^3} + [ \lambda + 2f ] & sym. \\
 -w\, v - \frac{\sqrt{G}\,r^2}{\gamma^4}
\left( \frac{\gamma^2 u}{r^2} \right)_\rho & -2\mu w &
\frac{\gamma_{\rho\rho}}{\gamma^3} + [ \lambda + f ]
+ \frac{4G u^2}{\gamma^2} \end{array} \right),
\nonumber\\
\bbS_2 &=& \gamma^2\left( \begin{array}{cccc}
 \frac{r_{\rho\rho}}{r\gamma^2} + \lambda
+ \frac{4G u^2}{\gamma^2} & sym. & sym. & sym. \\
 \sqrt{\lambda}\, v & [ \lambda + 2f ] + v^2 & sym. & sym. \\
 0 & \sqrt{2}\mu w & [ \lambda - 2 + 3f  ] & sym. \\
 \sqrt{2G}\frac{u_\rho}{\gamma^2} & -\sqrt{2\lambda}\, w & 0
& [ \mu^2 - f ] \end{array} \right),
\nonumber\\
\bbS_t &=& \left( \begin{array}{ccc}
 2\sqrt{\lambda}\,\gamma_\rho & 0 & 0 \\
 \frac{r}{\gamma}\left[ \frac{\gamma^2}{r} v \right]_\rho
+ 2\sqrt{G}\,\gamma w\, u &
 2\mu \gamma_\rho & -2(\gamma w)_\rho - 2\sqrt{G}\gamma\, uv \\
-\sqrt{2G}\,\mu\gamma\, u & 2\sqrt{2} (\gamma w)_\rho &
-\sqrt{2}\mu \gamma_\rho \\
-\sqrt{2G\lambda}\,\gamma\, u & 0 & \sqrt{2\lambda} \gamma_\rho
\end{array} \right),
\nonumber
\eea
where $\mu^2 = \ell(\ell+1)$ and $\lambda = \mu^2 - 2$ depend on the
angular momentum number $\ell$.
The constraint equations are found to be
\bea
0 &=& \frac{\gamma}{r}\,\p_\rho\!\left( \frac{r}{\gamma}\, h \right)
+ 2\sqrt{G}u\, b - \sqrt{\lambda}\gamma k
    - \gamma\, v\, c, \nonumber\\
0 &=& \gamma\,\p_\rho\!\left( \frac{1}{\gamma}\, a \right)
-\mu\gamma\, c - \sqrt{2}\gamma w\, d,
\nonumber \\
0 &=& \gamma\,\p_\rho\!\left( \frac{1}{\gamma}\, b \right)
+ \gamma w\, c + \frac{\mu}{\sqrt{2}}\gamma\, d
    - \sqrt{\frac{\lambda}{2}}\,\gamma\, e.
\label{Eq-WaveConsEYMLSphOdd}
\eea
How to solve the initial value problem in terms of gauge-invariant
quantities is discussed in Appendix \ref{App-B}.
A nice fact is that the cosmological constant does not appear
explicitly in the above system, it appears only via the
background quantities $S$, $N$ and $w$. As a consequence, the
stability can be discussed on the same lines as for
asymptotically flat solutions \cite{S-Diss}.
Below, we study perturbations which are smooth and which vanish
at the boundary points $\rho =\rho _{0}$
($r=0$ or $r=r_h$ as applicable)
and $\rho = \rho_{max}$ ($r=\infty $).
The spatial part of the wave operator defined in
(\ref{Eq-SymWaveEYMLOdd}) is symmetric for such perturbations, and
since the operator is real, self-adjoint extensions exist.
The spectrum of the self-adjoint extension which corresponds to
physical boundary conditions decides the linear stability of
the background solution.
It is, however, beyond the scope of this article to determine the
self-adjoint extensions. We will rather base our stability
argument on an energy estimate.

\subsection{The projection onto the constraint manifold and
stability}

In order to study the linear stability of the background
solutions, one has to consider the time evolution of perturbations
$U$, which are restricted to the constraint manifold defined by
(\ref{Eq-WaveConsEYMLSphOdd}). Our next aim is therefore to
separate the purely dynamical degrees of freedom from the
constraint violating modes.
Generalizing ideas introduced by Anderson {\it et al}.
\cite{AAL-CurvPert}, we define the constraint variables $u_c\,$
by the right-hand side of (\ref{Eq-WaveConsEYMLSphOdd}).
We are then looking for a dynamical variable, $u_p\,$, say,
such that the wave equation assumes the form
\be
\left[ \p_t^{\, 2} - \p_\rho^{\, 2}
+ \left( \begin{array}{cc} \bbV_{\! c} & 0 \\
\bbV_{\! pc} & \bbV_{\! p} \end{array} \right) \right]
\left( \begin{array}{c} u_c \\ u_p \end{array} \right) = 0,
\label{Eq-WaveConsDynSep}
\ee
when expressed in the new variables $(u_c, u_p)$.
The operators $\bbV_{\! c}\,$, $\bbV_{\! pc}$ and $\bbV_{\! p}$ are permitted to
have up to first order spatial derivatives only.
The components in $\bbV_{\! cp}$ have to vanish in order for the constraint
variables to evolve.
On the constraint manifold, $u_c = 0$, we then get
\bdm
\left[ \p_t^{\, 2} - \p_\rho^{\, 2} + \bbV_{\! p} \right] u_p = 0,
\edm
which is a wave equation for the dynamical variables $u_p\,$.
Of course, it would be nice if the transformation can be found
in such a way that the new potential $\bbV_{\! p}$ is automatically
symmetric. In our case, it turns out that this is possible.

As a simple example we first decouple the dynamical modes from
the constraint violating ones in the vacuum case \cite{SHB-PRD},
where the YM amplitudes $a$, $b$, $c$, $d$ and $e$ vanish, and where $w=1$.
The idea is to find a first order linear transformation of the form
\bdm
\left( \begin{array}{c} u_c \\ u_p \end{array} \right) =
\bbB\left( \begin{array}{c} h \\ k \end{array} \right), \;\;\;
\bbB = \p_\rho + \bbC,
\edm
such that the spatial part of the wave operator factorizes,
\be
-\p_\rho^{\, 2} + \left( \begin{array}{ll}
  \frac{r}{\gamma}\left( \frac{\gamma}{r} \right)_{\rho\rho}
+ \lambda\gamma^2 & 2\sqrt{\lambda}\gamma_{\rho} \\
2\sqrt{\lambda}\gamma_{\rho} & \frac{r_{\rho\rho}}{r} + \lambda\gamma^2
\end{array} \right) = \bbB^\dagger \bbB,
\label{Eq-WaveFactorize}
\ee
where $\bbB^\dagger = -\p_\rho + \bbC^T$ is the (formal) adjoint of $\bbB$.
The desired wave equation (\ref{Eq-WaveConsDynSep}) is then
obtained upon applying the operator $\bbB$ to the left of the original
wave equation (\ref{Eq-SymWaveEYMLOdd}). This yields
\bdm
\left[ \p_t^{\, 2} + \bbB\bbB^\dagger \right]
\left( \begin{array}{c} u_c \\ u_p \end{array} \right) = 0.
\edm
Since $\bbB\bbB^\dagger$ is symmetric, the fact that $\bbV_{\! cp}$
must vanish implies that $\bbV_{\! pc}$ has to vanish as well.
Provided that the factorization (\ref{Eq-WaveFactorize}) can be found,
the constraint and dynamical variables can therefore be decoupled
completely.

It is far from obvious that such a transformation exists,
since equation (\ref{Eq-WaveFactorize}) automatically implies that
the spatial operator is positive. Nevertheless, it turns out that
the factorization can be found in our case.
Indeed, by making the ansatz
\bdm
\bbB = \p_\rho + \left( \begin{array}{cc}
  \frac{\gamma}{r}\left(\frac{r}{\gamma}\right)_\rho
& -\sqrt{\lambda}\gamma \\
 -\sqrt{\lambda}\gamma & A  \end{array} \right),
\edm
where the first row of the matrix has been chosen such that $u_c$
is defined by the right-hand side of the first
equation in (\ref{Eq-WaveConsEYMLSphOdd})
and the second row such that there are no first order
derivatives in $\bbB^\dagger \bbB$,
equation (\ref{Eq-WaveFactorize}) is fulfilled if
$A = -\frac{r_\rho}{r}$.
The transformed spatial operator is
\bdm
\bbB\bbB^\dagger = -\p_\rho^{\, 2} + \left( \begin{array}{cc}
 \frac{\gamma}{r}\left( \frac{r}{\gamma} \right)_{\rho\rho}
+ \gamma^2\lambda & 0 \\
 0 & r\left( \frac{1}{r} \right)_{\rho\rho}
+ \gamma^2\lambda \end{array} \right).
\edm
Using $r(1/r)_{\rho\rho} = \gamma^2(2 - 6m/r)$ for a Schwarzschild
background, we recognize that the dynamical variables are governed by
the Regge-Wheeler equation \cite{RW}.
Since the spatial operator has the form $\bbB^\dagger\bbB$, it is
positive on the space of smooth perturbations with compact support.
Therefore, in the odd-parity sector, the linear stability of the
Schwarzschild black hole can be established without needing
the explicit form of the Regge-Wheeler potential.
This sort of topological argument will be important
below, since the background solutions in EYM theory are not
known in closed form.

We now show how to decouple the constraint and dynamical variables
for the full EYM wave equation (\ref{Eq-SymWaveEYMLOdd}).
Here, the ansatz is
\bdm
\left( \begin{array}{c} u_c \\ u_p \end{array} \right) = \bbB U, \;\;\;
\bbB = \p_\rho - \bbA + \bbC_S\, ,
\edm
where the antisymmetric matrix $\bbA$ is given above
and is introduced in order to reproduce the first order
derivatives in the wave operator in (\ref{Eq-SymWaveEYMLOdd}).
The matrix $\bbC_S$ is assumed to be symmetric. Writing this
matrix in block form,
\bdm
\bbC_S = \left( \begin{array}{cc} \bbC_1 + \bbA_1 & \bbC_t^T \\
\bbC_t & \bbC_2 \end{array} \right),
\edm
we must have
\bea
\bbC_1 & = & \left( \begin{array}{ccc}
  \frac{\gamma}{r} \left( \frac{r}{\gamma} \right)_\rho & 0 & 2\sqrt{G}\, u \\
  0 & -\frac{\gamma_\rho}{\gamma} & 0 \\
  0 & 0 & -\frac{\gamma_\rho}{\gamma}
\end{array} \right),
\nonumber \\
\bbC_t^T & = & \gamma\left( \begin{array}{cccc}
  -\sqrt{\lambda} &   -v &                    0 & 0 \\
                0 & -\mu &          -\sqrt{2} w & 0 \\
                0 &    w & \frac{\mu}{\sqrt{2}} & -\sqrt{\frac{\lambda}{2}}
\end{array} \right),
\nonumber
\eea
in order for the constraint variables to be defined as the
right-hand side of (\ref{Eq-WaveConsEYMLSphOdd}).
Hence, only the symmetric $4\times 4$ matrix $\bbC_2$ has to be matched.

Equating $\bbB^\dagger\bbB$ with the spatial part of the
wave operator,
$-\p_\rho^{\, 2} + \bbA\p_\rho + \p_\rho\bbA + \bbS$
yields the following equations.
First, we find the equation
\bdm
-\p_\rho\bbC_1 + \bbC_1^2 + \bbC_t^T\bbC_t + 2\bbA_1\bbC_1 =
\bbS_1 + \p_\rho\bbA_1\, ,
\edm
which is a consistency condition that can be shown to hold automatically.
Next, we have a linear algebraic equation for the unknown,
symmetric matrix $\bbC_2\,$,
\be
\bbC_t^T\bbC_2 = \bbQ^T,
\label{Eq-LEYMOddTransLinEq}
\ee
where $\bbQ \equiv \bbS_t + \p_\rho\bbC_t - \bbC_t\bbC_1 - \bbA_2\bbC_t\,$.
Explicitly, we find
\bdm
\bbQ^T = \gamma\left( \begin{array}{cccc}
  \sqrt{\lambda}\,\frac{r_\rho}{r} & 2\sqrt{G}\,w u &   -\sqrt{2G}\mu u & 0 \\
                                 0 &              0 & \sqrt{2}\, w_\rho & 0 \\
               \sqrt{G\lambda}\, u &        -w_\rho &                 0 & 0
\end{array} \right).
\edm
Finally, we get a differential equation for $\bbC_2$:
\be
-\p_\rho\bbC_2 + \bbC_2^2 + [\bbA_2, \bbC_2] = \bbT,
\label{Eq-LEYMOddTransDiffEq}
\ee
where $\bbT \equiv \bbS_2 - \bbC_t\bbC_t^T + \bbA_2^2\,$, so that
\bdm
\bbT = \left( \begin{array}{cccc}
  \frac{r_{\rho\rho}}{r} + 2G u^2 & sym. & sym. & sym. \\
  0 & \gamma^2 w^2 & sym. & sym. \\
  0 & -\frac{\mu}{\sqrt{2}}\, \gamma^2 w &
\gamma^2[ \frac{\lambda}{2} + w^2 ] & sym. \\
  \sqrt{2G}\, u_\rho & -\sqrt{\frac{\lambda}{2}}\, \gamma^2 w &
\frac{1}{2}\mu\sqrt{\lambda}\, \gamma^2 &
  \gamma^2[ \frac{\mu^2}{2} - w^2] - 2G u^2
\end{array} \right).
\edm
Equation (\ref{Eq-LEYMOddTransDiffEq}) can be shown to be consistent with
the linear equation (\ref{Eq-LEYMOddTransLinEq}).
More precisely, if we set ${\cal C} \equiv \bbC_2\bbC_t^T - \bbQ$,
we can show that the differential equation for $\bbC_2$ implies that
\bdm
-\p_\rho{\cal C} + (\bbC_2 + \bbA_2){\cal C} - {\cal C}(2\bbA_1 + \bbC_1) = 0.
\edm

Suppose that we found a solution to the equations
(\ref{Eq-LEYMOddTransLinEq}) and (\ref{Eq-LEYMOddTransDiffEq}).
Then, the wave equation (\ref{Eq-SymWaveEYMLOdd}) is equivalent to
the wave equation
\bdm
\left( \p_t^{\, 2} + \bbB\bbB^\dagger \right) V =
\left( \p_t^{\, 2} - \p_\rho^{\, 2} + \bbA\p_\rho
+ \p_\rho\bbA + \tilde{\bbS} \right) V = 0,
\edm
where $V = \bbB U$.
The symmetric potential $\tilde{\bbS}$ is given in
block form by
\bea
\tilde{\bbS}_1 &=& \bbS_1 + 2[\bbC_1,\bbA_1]
+ 2\p_\rho( \bbC_1 + \bbA_1), \nonumber\\
\tilde{\bbS}_t &=& \bbS_t + 2( \bbC_t\bbA_1 - \bbA_2\bbC_t)
+ 2\p_\rho\bbC_t, \nonumber\\
\tilde{\bbS}_2 &=& \bbS_2 + 2[\bbC_2,\bbA_2] + 2\p_\rho\bbC_2. \nonumber
\eea
Since $\tilde{\bbS}_1$ and $\tilde{\bbS}_t$ do not depend
on $\bbC_2\, $, they can be computed directly. One obtains
\bdm
\frac{\tilde{\bbS}_1}{\gamma^2} = \left( \begin{array}{ccc}
  \frac{1}{r\gamma}\left( \frac{r}{\gamma} \right)_{\rho\rho}
+ \lambda + v^2 + \frac{4Gu^2}{\gamma^2} &
    sym. & sym. \\
  \mu v & \frac{1}{\gamma}\left( \frac{1}{\gamma} \right)_{\rho\rho}
+ \lambda + 2f & sym. \\
 -w\, v + \sqrt{G}\left( \frac{u}{\gamma^2} \right)_\rho & -2\mu w &
  \frac{1}{\gamma}\left( \frac{1}{\gamma} \right)_{\rho\rho} + \lambda + f
\end{array} \right) ,
\edm
and $\tilde{\bbS}_t = 0$.
The fact that $\tilde{\bbS}_t$ vanishes shows that - provided
that we can solve the equation (\ref{Eq-LEYMOddTransDiffEq}) - the
constraint variable and the dynamical variables decouple from each
other and are both governed by a symmetric wave equation.
The evolution equation for the constraint variables is given by
the symmetric wave equation
\bdm
\left( \p_t^{\, 2} - \p_\rho^{\, 2} + \bbA_1\p_\rho + \p_\rho\bbA_1
+ \tilde{\bbS}_1 \right)u_c = 0,
\edm
and therefore, initial data which satisfies $u_c =\dot{u}_c = 0$
will satisfy the constraint equations for all later times.

If we have been able to factorize the spatial operator,
stability of our solutions is then automatic.
More precisely, one can show that for smooth perturbations
with compact support, the energy expression
\bdm
E = \frac{1}{2} \int\limits_{\rho_0}^{\rho_{max}}
\Big\{ (\dot{U}, \dot{U}) + (\bbB U, \bbB U) \Big\} \diff\rho ,
\edm
is constant in time.
Our boundary conditions are that all perturbations
vanish at the origin (or event horizon) and at infinity.
However, at this point, one can also require
less restrictive boundary conditions.
At the horizon, for example, it is sufficient
to require that $U$ and $\dot{U}$ are finite.
At infinity ($\rho = \rho_{max}$), one can impose
the outgoing wave condition $\dot{U} = -\bbB U$, which
means that there is no radiation coming from infinity.
This condition implies that the energy cannot increase in time.
Since $E \geq 0$, the kinetic energy must therefore be bounded
and exponentially growing modes cannot exist.
The remarks in Appendix \ref{App-B} then show that the metric and
the gauge potential themselves cannot grow exponentially.

\subsection{Factorization of the wave operator}

The result of the previous section is that
we have shown the stability of our solutions
once we have found a factorization of the spatial operator,
that is, a solution of the equations
(\ref{Eq-LEYMOddTransLinEq}) and (\ref{Eq-LEYMOddTransDiffEq}).
This is the subject of the present section.

For the moment, we exclude the Schwarzscild-adS and the RN-adS
backgrounds from the analysis below. These cases have to be
treated separately.
We start with the distinguished cases $\ell=0$ and $\ell=1$.
For $\ell=0$, only the YM amplitudes $a$ and $d$ are present.
The linear equation (\ref{Eq-LEYMOddTransLinEq}) then reduces to
$-\sqrt{2}\, w \bbC^{(\ell=0)}_2 = \sqrt{2}\, w_\rho$, with
the unique solution $\bbC^{(\ell=0)}_2 = -w_\rho/w$.
Equation (\ref{Eq-LEYMOddTransDiffEq}) is also fulfilled by this
choice.
Since we are interested in the fundamental solutions where $w$ has
no zeros, $\bbC^{(\ell=0)}_2$ is regular, and the transformed
potential turns out to be
\bdm
\tilde{\bbS}^{(\ell=0)}_2 =
  \gamma^2[ w^2 + 1 ] + 2\left( \frac{w_\rho}{w} \right)^2,
\edm
which is the same as the one obtained in \cite{W-Stable}
in the sphaleronic sector.
The potential
$\tilde{\bbS}^{(\ell=0)}_2$ is positive, and has the asymptotic
behaviour
\bea
&& \tilde{\bbS}^{(\ell=0)}_2 \longrightarrow \frac{2}{r^2} \;\;\;
   \hbox{as}\;\;\; r \longrightarrow 0, \nonumber\\
&& \tilde{\bbS}^{(\ell=0)}_2 \longrightarrow 0 \;\;\;
   \hbox{as}\;\;\; r \longrightarrow r_h, \nonumber\\
&& \tilde{\bbS}^{(\ell=0)}_2 \longrightarrow -\frac{\Lambda}{3}(1 + w_\infty^2)\;\;\;
   \hbox{as}\;\;\; r \longrightarrow \infty, \nonumber
\eea
the behaviour at $r=0$ being the same as for $p$-waves.
Since $\tilde{\bbS}^{(\ell=0)}_2$ is positive, linear stability
follows, as discussed at the end of the previous subsection.
[Note that for solutions where $w$ has zeros, $\bbC^{(\ell=0)}_2$ is
singular. In this case, the method presented in \cite{Volkov}
can be generalized and the existence of exactly $n$ unstable
modes established ($n$ being the number of nodes of
the function $w$).]

For $\ell=1$ the amplitudes $k$ and $e$ are absent.
The corresponding rows and columns in the matrices $\bbC_t$,
$\bbC_2$, $\bbS_t$, $\bbS_2$ and $\bbA_2$ have therefore to
be removed. [The consistency conditions checked so far remain
valid, since for $\ell=1$, the entries in the removed rows
and columns decouple from the entries in the other rows and columns.]
The linear equation (\ref{Eq-LEYMOddTransLinEq}) turns out to
have the unique solution
\bdm
\bbC_2^{(\ell=1)} = \frac{w_\rho}{1 - w^2}
\left( \begin{array}{rr} w & -1 \\ -1 & w \end{array} \right),
\edm
which also satisfies the differential equation (\ref{Eq-LEYMOddTransDiffEq}).
Thus, the constraint and dynamical variables decouple, and
the dynamical variables are governed by the following symmetric
wave equation:
\bdm
\left( \p_t^{\, 2} - \p_\rho^{\, 2} + \tilde{\bbS}_2^{(\ell=1)} \right)u_p = 0,
\edm
for
\bdm
u_p = \left( \p_\rho + \bbC_2^{(\ell=1)} \right) \left(
\begin{array}{c} c \\ e \end{array} \right)
    + \gamma\left( \begin{array}{ccc}
-2\sqrt{G}\,\frac{w^2-1}{r} & -\sqrt{2}     & w \\
                       0 & -\sqrt{2}\, w & 1
      \end{array} \right)
      \left( \begin{array}{c} h \\ a \\ b \end{array} \right),
\edm
where the symmetric potential is given by
\bdm
\tilde{\bbS}_2^{(\ell=1)} =
  \gamma^2\left( \begin{array}{cc} 2
+ 4G\frac{(w^2-1)^2}{r^2} & 4w \\ 4w & 1 + w^2 \end{array} \right)
 + \frac{2w_\rho^2}{(1-w^2)^2}
\left( \begin{array}{cc} 1 + w^2 & -2w \\ -2w & 1 + w^2 \end{array} \right).
\edm
For solutions in which $w$ has no zeros, the supersymmetric
transformation is regular provided that $w\neq \pm 1$.
However, as discussed in section \ref{Sect-2},
for solitonic solutions $w=\pm 1$ only at the origin,
and for black holes, $w^{2}$ is never unity unless
$w\equiv \pm 1$, in which case the geometry is Schwarzschild-adS.
The new potential is again positive and has a similar
asymptotic behaviour as that for $\ell=0$. Therefore, the stability
follows also in this case.

For solutions in which $w$ does have zeros, the transformation
is again regular provided that $w\neq \pm 1$.
In asymptotically flat space, $w$ does not cross $\pm 1$ for
all regular solitonic and black hole solutions, except
that $w= \pm 1$ at the origin for solitons \cite{breit}.
This is also the case for the majority of solitons and
black holes in adS, so we can conclude that these solutions
also have no unstable modes for $\ell =1$.
However, there are some regular solutions for which $w$
has zeros and does cross
$\pm 1$ away from the origin or event horizon \cite{W-Stable},
for which we are not able to draw conclusions about the existence
of unstable $\ell =1$ modes.

We now turn to the generic case $\ell\geq 2$:
The general solution to the linear equation (\ref{Eq-LEYMOddTransLinEq})
can be written into the form
\bdm
\bbC_2 = \bbD\left( \bbX_0 + T(\rho)\bbX_1 \right)\bbD,
\edm
where $\bbD = \diag(n_2,m_3, m_4,m_5)$ is a positive, constant
matrix (the values of $n_2$, ... , $m_5$, which are not relevant
for what follows, are given in Appendix \ref{App-A}).
The symmetric matrices $\bbX_0$ and $\bbX_1$ are
\bea
\bbX_0 &=& \left( \begin{array}{cccc}
  -2\lambda\mu^2\frac{r_\rho}{r} + 4\sqrt{G} f_1 w\, uv & sym. & sym. & sym. \\
  -2\sqrt{2G}\, f_1 w\, u & 2w w_\rho & sym. & sym. \\
   2\sqrt{2G}\, f_1 u & -2w_\rho & 0 & sym. \\
   4\sqrt{2G}\, w^2(1-w^2)u & 2w^2 w_\rho & -2w w_\rho & -f_2 w w_\rho
\end{array} \right), \nonumber\\
\bbX_1 &=& \left( \begin{array}{cccc}
   4w^2 v^2 & sym. & sym. & sym. \\
  -2\sqrt{2G}\, w^2 v & 2w^2 & sym. & sym. \\
   2\sqrt{2G}\, w v & -2w & 2 & sym. \\
  \sqrt{2}\, f_2 w v & -f_2 w & f_2 & \frac{1}{2} f_2^2
\end{array} \right), \nonumber
\eea
where $f_1 = \lambda + 2w^2$ and $f_2 = \mu^2 - 2w^2$.
So far, the function $T(\rho)$ is arbitrary.
Introducing this into equation (\ref{Eq-LEYMOddTransDiffEq})
yields the following non-linear first order differential equation
for $T$:
\be
-\p_\rho T + {\cal {A}} T^2 + {\cal {B}} T + {\cal {C}} = 0,
\label{Eq-KeyEq}
\ee
with
\bea
\mu^2\lambda {\cal {A}} &=& \frac{8G}{r^2} w^2(1-w^2)^2
+ 4\left( w^2 - 1 - \frac{\lambda}{4} \right)^2 + 4\lambda
+ \frac{7}{4}\lambda^2, \nonumber\\
\mu^2\lambda {\cal {B}} &=& 8\left[ \frac{2G}{r^2}(\lambda + 2w^2)
+ 1 \right] (w^2-1)w w_\rho\, ,
\nonumber \\
\mu^2\lambda {\cal {C}} &=& \left[ \frac{2G}{r^2}(\lambda + 2w^2)
+ 2\lambda + 4w^2 \right] w_\rho^2
- \mu^2\lambda\left( \frac{\lambda}{2} + w^2 \right)\gamma^2.
\label{eq-ABC}
\eea
To summarize, it is sufficient to show that the single
differential equation (\ref{Eq-KeyEq}) admits a global
solution with appropriate boundary conditions in
order to show the stability of the evolution system
(\ref{Eq-SymWaveEYMLOdd}), which is a wave equation for
seven amplitudes.
In the next subsection we shall complete the stability
proof by showing the existence of a globally regular
solution to (\ref{Eq-KeyEq}).

Finally, we turn to the stability of the Schwarzschild-adS and
RN-adS black holes.
The stability of the Schwarzschild-adS metric can be established
on the same lines as above, but one has to take into account that
for $\ell=1$, the matrix $\bbC_2$ is no longer uniquely specified
by the linear equation (\ref{Eq-LEYMOddTransLinEq}), and one
obtains a differential equation similar to (\ref{Eq-KeyEq}).
Alternatively, one can also use the perturbation formalism
presented in Ref. \cite{SHB-Odd}, since in the odd-parity sector,
the cosmological constant does not appear explicitly.

While the stability proof goes through for $\ell\geq 1$,
the RN-adS solution turns out to be unstable with
respect to odd-parity radial perturbations of the non-Abelian part
of the YM field which is parametrized by the amplitude $d$.
When $w\equiv 0$, the amplitudes $a$ and $d$ decouple, and
$d$ is governed by the equation
\bdm
\left( \p_t^{2} -\p_{\rho }^{2}-\frac {N}{r^{2}} \right) d =0.
\edm
An easy way to find the instability is to use a trial function \cite{VG-trial}.
We define a sequence of functions $d(\rho )$ as follows:
\bdm
d(\rho ) = {\cal {Z}} \left( \frac {\rho }{J} \right),
\edm
for $J=1,2,\ldots $.
The function ${\cal {Z}}(\rho )\in [0,1]$ is defined by
\cite{VG-trial}
\bea
{\cal {Z}} (\rho ) & = & 0 \qquad
{\mbox {for $\rho < -P-1$,}}
\nonumber \\
{\cal {Z}}(\rho ) & = & 1 \qquad
{\mbox {for $\rho > -P$,}}
\nonumber
\eea
and
\bdm
0\le \frac {d{\cal {Z}}}{d\rho } \le Q
\qquad {\mbox {for $\rho \in [-P-1,-P]$,}}
\edm
where $P$ and $Q$ are arbitrary positive numbers.
The expectation value of the spatial operator yields,
after some calculation,
\bea
\left\langle\, d\, \left| -\p_\rho{\, ^2}
- \frac{N}{r^2}\, \right|\, d\, \right\rangle & = &
\int\limits_{r_h}^{\infty} \left( N d_r^{\, 2} - \frac{d^2}{r^2}
\right) \diff r
\nonumber \\
& \le &
\frac {Q^{2}}{J} -\frac {1}{r_{h}},
\nonumber
\eea
while $\langle\, d\, | \, d\, \rangle < \infty$.
This expectation value is negative
provided we choose $J$ to be sufficiently large.
Therefore we conclude that there are exponentially growing modes
and the RN-adS black hole is unstable.

\subsection{Global solutions to equation (\ref{Eq-KeyEq})}

It remains at this stage to show that equation
(\ref{Eq-KeyEq}) has globally regular solutions.
We shall begin by considering only those solitons
and black holes for which $w$ has no zeros,
since these are the solutions of most interest
to us.

In order to see that equation (\ref{Eq-KeyEq})
admits global solutions, one rewrites it as a linear,
second order differential equation using the transformation
\be
T = -\frac{1}{{\cal {A}}}\frac{z_\rho}{z}.
\label{eq-Tdef}
\ee
Going back to the radial coordinate $r$, this yields
\be
\frac {\p ^{2}z}{\p r^{2}} +\left[
\frac {1}{NS} \frac {\p (NS)}{\p r} - {\tilde {{\cal {B}}}}
-\frac {1}{{\cal {A}}} \frac {\p {\cal {A}}}{\p r} \right]
\frac {\p z}{\p r} + {\cal {A}} {\tilde {{\cal {C}}}}z=0,
\label{eq-z}
\ee
where
\bea
\mu ^{2} \lambda {\tilde {{\cal {B}}}} & = &
8 \left[ \frac {2G}{r^2} (\lambda + 2w^{2}) +1 \right]
(w^{2} -1) w w_{r} ,
\nonumber \\
\mu ^{2} \lambda {\tilde {{\cal {C}}}} & = &
\left[ \frac {2G}{r^{2}} (\lambda + 2w^{2}) + 2\lambda
+4w^{2} \right] w_{r}^{2}
-\frac {\mu ^{2}\lambda }{Nr^{2}} \left( \frac {\lambda }{2} + w^{2}
\right) .
\label{eq-BC}
\eea
The equation (\ref{eq-z}) is regular everywhere except at the
origin, infinity, and (if one exists) the event horizon $r=r_{h}$,
where there are regular singular points.
Using the asymptotic expansions and the standard Frobenius method,
the solutions for $z$ near each of these points can be found.

Near the origin, for globally regular (solitonic) solutions,
the functions in (\ref{eq-z}) behave as:
\bea
\mu ^{2}\lambda  {\cal {A}} & = & 2\lambda ^{2} + 4 \lambda
+O(r^{2}) ,
\nonumber \\
\mu ^{2}\lambda  {\tilde {{\cal {B}}}} & = & O(r^{3}) ,
\nonumber \\
\mu ^{2} \lambda {\tilde {{\cal {C}}}} & = &
-\frac {\mu ^{2}\lambda }{r^{2}}
\left( \frac {\lambda }{2} +1 \right) +O(1) , \nonumber
\eea
so that the origin is a regular singular point, and
the two linearly independent solutions have the form
\bdm
z  \sim r^{\ell +1}, \quad r^{-\ell }  .
\edm
If, instead, we are considering black hole solutions with
an event horizon at $r=r_{h}$, then
${\cal {A}}$ and ${\tilde {{\cal {B}}}}$ are
regular functions at the event horizon, while
${\tilde {{\cal {C}}}}$ diverges
as $(r-r_{h})^{-1}$.
Therefore, again there is a regular singular point, this
time the indicial equation for $z$ has a repeated root of zero,
so that the linearly independent solutions have the form:
\bdm
z = O(1), \quad O(\log (r-r_{h})).
\edm
As expected, the negative cosmological constant means that
the analysis at infinity is different to that in the
asymptotically flat case \cite{S-Diss}.
Letting $s=1/r$, the behaviour of the functions in (\ref{eq-z})
is:
\bea
N S & = & -\frac{\Lambda}{3s^2} + O(1/s),
\nonumber\\
{\cal {A}} & = & A_{\infty } + A_{1}s +O(s^{2}),
\nonumber \\
{\tilde {{\cal {B}}}} & = & B_{\infty }s^{2} +O(s^{3}),
\nonumber \\
{\tilde {{\cal {C}}}} & = & C_{\infty }s^{4} +O(s^{5}),
\nonumber
\eea
which implies that the point $s=0$ is, in fact, a regular point.
As a consequence, the asymptotic forms of the linearly independent
solutions for $z$ at infinity are:
\bdm
z=O(1), \quad O(1/r).
\edm

The proof of the existence of global solutions to equation
(\ref{eq-z}) can now proceed as in the asymptotically flat case
\cite{S-Diss}.
Consider firstly black hole solutions.
We start near the event horizon with the solution having $z = O(1)$
(i.e. with no logarithmic term),
so that $z$ is positive for sufficiently small $r-r_{h}$.
At a maximum of $z$, where $dz/dr=0$, from (\ref{eq-z}) we have
\bdm
\frac {d^{2}z}{dr^{2}} = -{\cal {A}}{\tilde {{\cal {C}}}} z .
\edm
It is clear from (\ref{eq-ABC}) that ${\cal {A}}>0$ always, so we need
to investigate the sign of ${\tilde {{\cal {C}}}}$.
In the asymptotically flat case \cite{S-Diss}, it was checked
numerically that ${\tilde {{\cal {C}}}}<0$, but in our situation, we
can prove this analytically for sufficiently large
$|\Lambda |$.

First fix $\ell =2$.
It was proved in \cite{W-Stable} (see also section
\ref{Sect-2}) that, as
$|\Lambda | \rightarrow \infty $,
\bdm
N\sim O(|\Lambda |), \qquad
\frac {dw}{dr} \sim  o \left( |\Lambda |^{\frac {1}{2}} \right)
\edm
for all $r$.
Hence, for sufficiently large $|\Lambda |$, both terms on
the right-hand-side of (\ref{eq-BC}) are vanishing,
but the first term is $o(|\Lambda |^{-1})$ and the second
term $O(|\Lambda |^{-1})$, so the first term is vanishing
more rapidly.
Therefore, for sufficiently large $|\Lambda |$, we have
that ${\tilde {{\cal {C}}}}<0$ for all $r$, for this particular value
of $\ell$.
If we now replace $\ell \rightarrow \ell +1$, then
$\mu ^{2} \lambda {\tilde {{\cal {C}}}} \rightarrow \mu ^{2} \lambda
{\tilde {{\cal {C}}}}+{\cal {F}}$, where a lengthy calculation
(using the fact that ${\tilde {{\cal {C}}}}<0$ for the original
value of $\ell $) yields
\bea
{\cal {F}} & < & \frac {\lambda + 2 w^{2}}{(\ell -1)^{2}(\ell +2)}
\left[ -\frac {2G\lambda }{r^{2}} (2\ell ^{2}+6\ell )
-2\lambda (2\ell ^{2}+6\ell )
\right. \nonumber \\
& & \left.
-8w^{2}(\ell -1)(\ell +2) \left(
\frac {G}{r^{2}}+1 \right) \right]
\nonumber \\ & < & 0,
\nonumber
\eea
where $\lambda $ and $\mu $ are calculated with the original
value of $\ell $.
Therefore, as $\ell $ increases ${\tilde {{\cal {C}}}}$ decreases,
and is therefore negative for all values of $\ell $, for
sufficiently large $|\Lambda |$.

\begin{figure}
  \begin{center}
    \includegraphics[height=0.35\textheight]{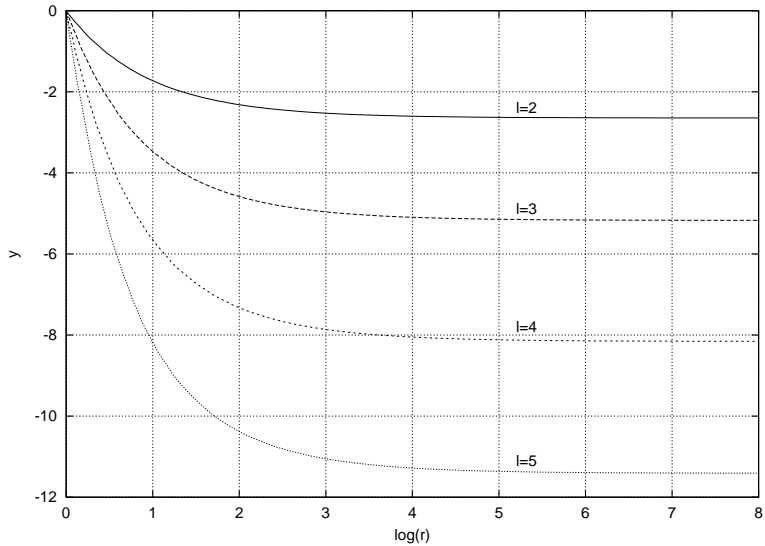}
    \caption{The function $y\equiv T/S$ for $\ell=2,3,4,5$.
             The background solution is a black hole with
             $\Lambda=-100$, $r_h=1$, and $w(r_h) = 0.9$.}
  \label{Fig-1}
  \end{center}
\end{figure}

We conclude that $z$ cannot have a maximum if it is
positive, nor minimum if it is negative.
Therefore, since $z$ is positive close to the event horizon,
it will remain strictly positive, and take the form
\bdm
z = z_{\infty } + O(r^{-1})
\edm
at infinity, where $z_{\infty} > 0$.
Since ${\cal {A}}$ is positive, and $z$ has no zeros, using the
definition (\ref{eq-Tdef}) we see that $T$ is regular
and exists globally with the asymptotic behaviour
\bdm
T\sim (r-r_{h})
\edm
near the event horizon, and
\bdm
T = O(1)
\edm
at infinity.
As an example, the function $T/S$ for a particular black hole
solution with $\Lambda= -100$ is shown in figure \ref{Fig-1}.

For globally regular solutions, a similar argument holds,
using the solution $z\sim r^{\ell +1}$ near the origin.
In this case the function $T$ behaves like $r^{-1}$ for
small $r$ but the behaviour at infinity is unchanged.

A careful analysis reveals that again, the behaviour of the
transformed potential $\tilde{\bbS}_2$ is similar to the $\ell=0$
case: $\tilde{\bbS}_2 = O(1/r^2), O(r - r_h), O(1)$ near $r=0, r_h,
\infty$, respectively.

We have now completed the proof that those black holes
and solitons which have no unstable modes in the
odd-parity sector for spherically symmetric perturbations
also have no unstable modes in the odd-parity sector
when we consider non-spherically symmetric perturbations.
In other words, we have proved the (odd-parity) stability
of those solutions in which $w$ has no zeros and $|\Lambda |$
is sufficiently large.

A similar analysis shows that the Bartnik-McKinnon solitons and
the corresponding black holes with hair have no unstable modes
with $\ell > 0$ and odd parity \cite{S-Diss}. The only
difference there, is that the background quantities imply that
the function $T$ behaves like $1/r$ near infinity. Instead of
approaching a constant value, the transformed potential falls off
as $1/r^2$ when $r\rightarrow\infty$.

The question, therefore, is whether our analysis can be
extended to other values of $|\Lambda |$.
The answer, surprisingly, is no.
The crucial part of the work is the sign of the
function ${\tilde {\cal {C}}}$.
For large $|\Lambda |$ we have been able to show analytically
that ${\tilde {\cal {C}}}$ is always negative.
When $\Lambda =0$, numerical analysis showed this to also be
the case \cite{S-Diss}.
However, for small, negative $\Lambda $, we have found
numerically that ${\tilde {\cal {C}}}$ can be positive
and the solutions to (\ref{Eq-KeyEq}) blows up
(for example, for black hole solutions with $\Lambda =-0.001$,
$r_{h}=1$ and $w(r_{h})=0.642$).
Therefore our analysis does not extend and it may be that
these solutions have additional instabilities (they are already
unstable for spherically symmetric perturbations).
In order to find such instabilities, one could use
the nodal theorem for coupled, symmetric systems of
Schr\"odinger equations proved in Ref. \cite{AQ-Nodal}
to count the number of bound states of the spatial operator.

\section{Conclusions}
In this article we have studied the linear stability
of soliton and hairy black hole solutions of
${\mathfrak {su}}(2)$ Einstein-Yang-Mills theory
with a negative cosmological constant.
We have applied a recently developed perturbation formalism
which is based on curvature quantities.
The advantage of this approach is that one can work
with gauge-invariant quantities, which are governed by a
wave equation whose spatial part is symmetric.
As a consequence, we were able to decouple the constraint
variables by a supersymmetric-like transformation, and to show
analytically the linear stability of soliton and black holes with
respect to odd-parity fluctuations.
We stress that such a proof is unlikely to exist in a
gauge-invariant metric formulation, since when non-Abelian
gauge fields are coupled to the metric, the metric approach
fails to yield a symmetric wave operator in a natural way.

Our main result concerns those solutions in which the
function $w$ which determines the gauge field potential
has no zeros and the cosmological constant is large
and negative.
These solutions are of particular interest as it was
already known that they have no modes of
instability under spherically symmetric perturbations.
We have proved that this holds also for general, linear,
perturbations in the odd-parity sector.
This result is significant because it shows that these
static configurations possess no topological instabilities.
Therefore the presence of a sufficiently large (negative) cosmological
constant (which means that there is a large
gravitational potential away from the black hole horizon, or
the centre of the solitons) stabilizes the situation.
This may be understood heuristically by analogy with the
situation in quantum field theory in curved space.
A black hole can be in thermal equilibrium with a bath
of radiation at the Hawking temperature,
however in asymptotically flat space this
equilibrium is unstable.
Stability can be restored either by placing the whole system
in a box, or if the black hole instead resides in
asymptotically anti-de Sitter space \cite{HP}, provided
the cosmological constant is sufficiently large and
negative.
In some sense, then, the structure of asymptotically
adS space plays a similar role in our situation to
placing the black hole or soliton in a box, which
means that the gauge field cannot escape to infinity,
as happens when the corresponding configurations
in asymptotically flat space are perturbed \cite{zhou}.
Further, because of the above stability of the Hartle-Hawking
state, the black holes we have studied in this paper are
precisely those which are most of interest from the
quantum field theory point of view.
Therefore it is important to establish their classical
stability before studying the properties of quantum
fields propagating on these geometries.
This will be the subject of future work.

Classically, some open questions remain.
Firstly, we need to investigate the stability
of these solutions under non-spherically symmetric
even-parity perturbations.
The even-parity is generally
less amenable to analysis,
since its properties depend crucially on
the detailed structure of the equilibrium
configurations.
However, in the situation in which we are interested,
when the cosmological constant is large and negative,
it was found in \cite{W-Stable} that, for
spherically symmetric perturbations,
the even-parity sector simplified considerably
and stability was proved analytically in this case.
Therefore it seems reasonable to suppose that
simplification may be possible also
for non-spherically symmetric perturbations.

Next, we have not examined the zero modes of the
pulsation equations, i.e. the stationary solutions to
the perturbation equations. Similarly to the asymptotically
flat case \cite{BHVS-Rotation}, we except to find slowly rotating and
charged solitons and black holes, the latter ones generalizing
the Kerr-adS metric.

We shall return to these questions in a
subsequent publication.

\section*{Acknowledgments}
O.S. would like to thank H. Morales-T\'ecotl and M. Tiglio for helpful
discussions.
This work was supported in part
by the Swiss National Science Foundation.

\appendix
\section{Harmonic decomposition of the wave operator}
\label{App-A}

With respect to the background 3-metric
\bdm
\gbartens = \frac{\diff r^2}{N} + r^2 \diff\Omega^2,
\edm
odd-parity perturbations of the extrinsic curvature can be
expanded according to
\be
L_{rr} = 0, \;\;\;
L_{rB} = \frac{ \tilde{h} }{\sqrt{N}}\, S_B, \;\;\;
L_{AB} = r\tilde{k}\, 2\nabhat_{(A} S_{B)}\, .
\label{Eq-ExpLij}
\ee
Here and in the following, capital indices refer to coordinates
on the 2-sphere. In the above equation,
$S_B = S^{\ell m}_B$ are the transverse vector
harmonics, and can be expressed in terms of the standard spherical
harmonics $Y \equiv Y^{\ell m}$ as $S_B = (\asthat\diff Y)_B$,
where a hat refers to the standard metric on the 2-sphere.

Similarly, the electric YM field is expanded into
${\mathfrak {su}}(2)$-valued vector harmonics with odd parity
(see \cite{SHB-Odd} for details),
\bea
{\cal E}_r &=& \frac{\tilde{a}}{r\sqrt{N}}\, X_1
+ \frac{\tilde{b}}{r\sqrt{N}}\, X_2,
\nonumber \\
{\cal E}_A &=& \tilde{c}\, \taur Y_A + \tilde{d}\, Y \tau_A +
\tilde{e}\left( \nabhat_A X_2 + \frac{1}{2}\mu^2 Y \tau_A\right),
\label{Eq-ExpEi}
\eea
where in terms of the Pauli matrices $\ul{\sigma} = (\sigma^i)$,
$\taur = \ul{e}_r\cdot\ul{\sigma}/(2i)$,
$\tau_A = \ul{e}_A\cdot\ul{\sigma}/(2i)$.
Here $X_1$ and $X_2$ are the
${\mathfrak {su}}(2)$-valued harmonics
\bdm
X_1 = Y \, \taur , \;\;\;
X_2 = \ghat^{AB} \nabhat_A Y \tau_B\, .
\edm
The tensor harmonics in the expansions (\ref{Eq-ExpLij}) and
(\ref{Eq-ExpEi}) are chosen to be orthogonal. After the rescaling
\bdm
\tilde{h} = n_1 h, \;\;\;
\tilde{k} = n_2 k, \;\;\;
\tilde{a} = m_1 a, \;\;\;
\tilde{b} = m_1 b, \;\;\;
\tilde{c} = m_1 c, \;\;\;
\tilde{d} = m_1 d, \;\;\;
\tilde{e} = m_1 e,
\edm
where in terms of $\mu^2 = \ell(\ell+1)$ and $\lambda = \mu^2 -2$,
the coefficients are given by
\bea
&& n_1 = \frac{1}{\sqrt{2\mu^2}}\, , \;\;\;
   n_2 = \frac{1}{\sqrt{2\mu^2\lambda}}\, , \nonumber\\
&& m_1 = \frac{1}{\sqrt{2G}}\, , \;\;\;
   m_2 = m_3 = \frac{1}{\sqrt{2G\mu^2}}\, , \;\;\;
   m_4 = \frac{1}{\sqrt{4G}}\, , \;\;\;
   m_5 = \frac{1}{\sqrt{G\mu^2\lambda}}\, , \nonumber
\eea
the expansion is normalized such that
\bdm
\sprod{ (L, {\cal E}) }{ (L, {\cal E}) }
= \int (h^2 + k^2 + a^2 + b^2 + c^2 + d^2 + e^2) \frac{\diff r}{\sqrt{N}}\, ,
\edm
where $\sprod{.}{.\, }$ denotes the scalar product defined
by (\ref{Eq-InnerProduct}).
Note that for $\ell=1$, $\nabhat_{(A} S_{B)}$ and
$\nabhat_A X_2 + \frac{1}{2}\mu^2 Y \tau_A$ vanish, and therefore,
the amplitudes $k$ and $e$ do not exist in those cases.
For $\ell=0$, the function
$Y$ is constant and only the amplitudes $a$ and $d$
are present.

An efficient way to perform the harmonic decomposition of the wave
operator is to compute the energy expression which
corresponds to $\hat{\Lambda}_{ij}$ and $\Lambda^{(YM)}$,
\bdm
E = E_{grav} + E_{YM} + E_{int},
\edm
where
\bea
E_{grav} &=& \frac{1}{2} \int\left(
     \frac{1}{\alpha^2} \dot{L}^{ij}\cdot\dot{L}_{ij}
   + \nabbar^k L^{ij}\cdot\nabbar_k L_{ij}
   + 2L^{ij} \bar{R}^k_{\; i} L_{jk}
   - 2L^{ij} \bar{R}_{kilj} L^{kl} \right. \nonumber\\
  && \left. +\, \frac{8}{\alpha} L^{ij}\nabbar_i \left( \alpha^k L_{jk} \right)
   - 2L^{ij}\nabbar^k \left( \frac{\alpha_i}{\alpha} \right) L_{jk}
   - \frac{L^{ij}}{\alpha^2} \nabbar_i\alpha^2\nabbar_j
   \left( \frac{A}{\alpha} \right) \right. \nonumber\\
  && \left. - 2\Lambda L^{ij} L_{ij}
   + 4G\Trace\left\{ L^{ij}\bar{F}^k_{\;\, i} \bar{F}^l_{\; j} L_{kl}
   + \frac{1}{4} \bar{F}_{kl} \bar{F}^{kl} L_{ij} L^{ij}
   \right\}\right) \alpha\sqrt{\gbar}\,\diff\,x^3, \nonumber\\
E_{YM} &=& G \int\Trace\left\{ \frac{1}{\alpha^2} \dot{\cal E}^i\dot{\cal E}_i
 + 2\Diffbar^{[i}{\cal E}^{j]}\cdot \Diffbar_{[i}{\cal E}_{j]}
 + \alpha^2 \left[ \Diffbar^j\left(
     \frac{{\cal E}_j}{\alpha} \right) \right]^2 \right. \nonumber\\
&& \left. \qquad\qquad +\, \bar{F}^{ij}[{\cal E}_i, {\cal E}_j]
 + 4G\Trace( \bar{F}^l_{\; k} {\cal E}_l )
 \bar{F}^{ik}{\cal E}_i \right\} \alpha\sqrt{\gbar}\,\diff\,x^3, \nonumber\\
E_{int} &=& -4G\int\Trace\left\{
  L^{ij}\bar{F}_i^{\; k}\Diffbar_k {\cal E}_j
 + \frac{1}{\alpha^2} L^{ij} {\cal E}_k
 \Diffbar_i \left( \alpha^2\bar{F}_j^{\; k} \right) \right. \nonumber\\
&& \left. \qquad\qquad +\, \frac{1}{2} {\cal E}^i \bar{F}_{ij}
 \nabbar^j \left( \frac{A}{\alpha} \right) \right\} \alpha\sqrt{\gbar}\,\diff\,x^3.
\nonumber
\eea
For a spherically symmetric background with no electric field,
non-vanishing background quantities are given by
\bea
& \bar{R}^r_{\; ArB} = -\frac{r}{2}N_r \ghat_{AB},
& \bar{R}^D_{\; CAB} = 2(1 - N) \delta^D_{\; [A} \ghat_{B]C},
\nonumber\\
& \bar{R}_{rr} = -\frac{N_r}{rN},
& \bar{R}_{AB} = \left( 1 - N - \frac{r}{2}N_r \right)\ghat_{AB},
\nonumber\\
& \bar{F}_{rB} = -w_r\, \hat{\vep}^A_{\; B} \tau_A,
& \bar{F}_{AB} = (w^2 - 1)\taur\, \hat{\vep}_{AB}.
\nonumber
\eea
Using this, the background equations and the expansions (\ref{Eq-ExpLij})
and (\ref{Eq-ExpEi}), the energy expression yields, after
integrating over the spherical variables,
\bdm
E = \frac{1}{2} \int \Big\{ (\dot{U}, \dot{U}) + (\p_\rho U, \p_\rho U)
  + 2(U, \bbA \p_\rho U) + (U, \bbS U) \Big\} \diff\rho,
\edm
where $U = (h,a,b,k,c,d,e)$, and where $\bbA$ and $\bbS$ are given
in section \ref{Sect-4}. The radial coordinate $\rho$ is defined by
$\p_\rho = SN\p_r$, and $(.\, ,.)$ denotes the standard scalar product.
The wave equation (\ref{Eq-SymWaveEYMLOdd}) follows directly from
this energy functional.

\section{The initial value formulation}
\label{App-B}

In this appendix, we give a gauge-invariant
formulation of the initial value problem.
In order to solve the linearized EYM equations, we
have to take into account the Hamiltonian constraint and all evolution
equations which we had differentiated in time.
These equations cannot be described in terms of the
linearized extrinsic curvature and the electric YM field alone,
since perturbations of the 3-metric and the magnetic gauge
potential appear (and not only their time derivatives).
However, if the background is spherically symmetric,
one can give a formulation in terms of gauge-invariant quantities
which we introduced in an earlier article \cite{SHB-Odd}.

It turns out that the ``missing'' constraint equations (i.e. the
linearized Hamiltonian constraint and the relevant evolution equations
we had differentiated in time) are equivalent to the $\rho$-components
of the equations (41), (42) and (43),
and to equation (46) of Ref. \cite{SHB-Odd}
[The presence of the cosmological constant does not modify
those perturbation equations.]
Using the gauge-invariant amplitudes $H$, $A$, $B$ and $C$
defined there, these equations become
\bea
2\p_t\tilde{h} &  = & -\lambda\frac{\gamma}{r} H_\rho
    + \frac{4G\gamma}{r} \left[
      (w^2-1)(A_\rho - w B_\rho) - w_\rho C - \frac{(1 - w^2)^2}{r^2} H_\rho \right],
\nonumber\\
 \p_t\tilde{a}&  = &
   \gamma\left[ (\mu^2 + 2w^2) A_\rho - 2\mu^2 w B_\rho - 2w C_\rho + 2w_\rho C
   + \mu^2(1 - w^2)\frac{H_\rho}{r^2} \right],
\nonumber\\
 \p_t\tilde{b} & = & \gamma\left[
   - 2w A_\rho + (\mu^2 + w^2 - 1) B_\rho + C_\rho
   - w(1 - w^2)\frac{H_\rho}{r^2} \right],
\nonumber \\
 \p_t\left( \tilde{d} + \frac{1}{2}\mu^2\,\tilde{e} \right)
\hspace{-0.5in} & &  \nonumber\\
 & = & w\p_\rho A_\rho + 2w_\rho A_\rho + \gamma^2[ \mu^2 + w^2 - 1 ]C - C_{\rho\rho}
   + \mu^2 w_\rho \frac{H_\rho}{r^2}\, .
\label{Eq-DottedEvolEq}
\eea
Furthermore, one has the relation between the gauge-invariant
amplitudes of Ref. \cite{SHB-Odd} and the curvature-based
amplitudes $\tilde{h}$, $\tilde{k}$, $\tilde{a}$, ... introduced
in appendix A of this article. For $\ell\geq 2$, one has
\bea
& H_t = -2r\tilde{k}, &
  \dot{H}_\rho = 2\alpha\tilde{h} - 2r^2\p_\rho\left(
     \frac{\tilde{k}}{r} \right),
\nonumber\\
& A_t = \left( \tilde{c} + w\tilde{e} - 2\frac{w^2-1}{r}\tilde{k} \right), &
   \dot{A}_\rho = -\gamma\tilde{a} + \p_\rho A_t\, ,
\nonumber\\
& B_t = \tilde{e}, &
  \dot{B}_\rho = -\gamma\tilde{b} + \p_\rho\tilde{e} - 2\frac{w_\rho}{r}\tilde{k},
\nonumber \\
& \dot{C} = -\tilde{d} - \frac{1}{2}\mu^2\tilde{e} + w A_t\, . &
\label{Eq-EYMOldNewAmpl}
\eea
The initial value problem for $\ell\geq 2$ can be solved as follows.
First, we choose any functions $H_\rho = H_\rho^{(0)}$,
$A_\rho = A_\rho^{(0)}$, $B_\rho = B_\rho^{(0)}$ and
$C = C^{(0)}$ on an initial time slice, $\Sigma_{t=0}$, say.
Next, one solves the momentum constraint equations (\ref{Eq-WaveConsEYMLSphOdd})
for $U = (h,a,b,k,c,d,e)$ on the initial time slice. A convenient
way to do this is to freely specify the functions $c$, $d$, $e$ and $k$
and to compute $h$, $a$ and $b$ using (\ref{Eq-WaveConsEYMLSphOdd}).
Then, the time derivative of $U$ on $\Sigma_0$ is consistently
given by equations (\ref{Eq-DottedEvolEq}) and the time derivatives
of the momentum constraint equations. The amplitudes $U$ are then
evolved using the symmetric wave equation (\ref{Eq-SymWaveEYMLOdd}).
Finally, the gauge-invariant amplitudes $H$, $A$, $B$ and $C$,
parametrizing the metric and the gauge potential are obtained
from (\ref{Eq-EYMOldNewAmpl}) after integration over $t$, where
the integration ``constants'' are given by
$H_\rho^{(0)}$, $A_\rho^{(0)}$, $B_\rho^{(0)}$ and $C^{(0)}$.
For $\ell=1$, the perturbation equations can be solved
in a similar manner, but one has
to take into account that $C = 0$ and that the relation
(\ref{Eq-EYMOldNewAmpl}) is different in that case.

Since we have shown that the pulsation equations admit no
solutions which grow exponentially in time when $|\Lambda|$ is
large enough, the relations (\ref{Eq-EYMOldNewAmpl})
show that the same must hold for the gauge-invariant quantities
$H$, $A$, $B$, and $C$, parametrizing the metric and the
gauge-potential. This completes our stability analysis.


\begin{thebibliography}{10}

\bibitem{bartnik}
R. Bartnik and J. MacKinnon,
Phys. Rev. Lett. {\bf {61}}, 141 (1988).

\bibitem{bizon}
M.S. Volkov and D.V. Gal'tsov,
JETP Lett. {\bf 50}, 346 (1989);
\newline
H.P. K\"unzle and A.K.M. Masood-ul-Alam,
J. Math. Phys. {\bf 31}, 928 (1990);
\newline
P. Bizon,
Phys. Rev. Lett. {\bf 64}, 2844 (1990).

\bibitem{review}
M.S. Volkov and D.V. Gal'tsov,
Phys. Reps. {\bf {319}}, 2 (1999).

\bibitem{brod2}
O. Brodbeck and N. Straumann,
J. Math. Phys. {\bf {37}}, 1414 (1996).

\bibitem{galtsov}
D.V. Gal'tsov and M.S. Volkov,
Phys. Lett. {\bf {B273}}, 255 (1991).

\bibitem{brod1}
O. Brodbeck, M. Heusler, G. Lavrelashvili, N. Straumann
and M.S. Volkov,
Phys. Rev. {\bf {D54}}, 7338 (1996).

\bibitem{bjork}
J. Bjoraker and Y. Hosotani,
Phys. Rev. Lett. {\bf {84}}, 1853 (2000);
\newline
J. Bjoraker and Y. Hosotani,
Phys. Rev. {\bf {D62}}, 043513 (2000).

\bibitem{W-Stable}
E. Winstanley,
Class. Quantum Grav. {\bf 16}, 1963 (1999).

\bibitem{S-Diss}
O. Sarbach,
{\em On the generalization of the Regge-Wheeler equation for
self-gravitating matter fields},
PhD thesis (unpublished).

\bibitem{baacke}
J. Baacke and H. Lange,
Mod. Phys. Lett. {\bf {A7}}, 1455 (1992).

\bibitem{BHS-Letter}
O. Brodbeck, M. Heusler, and O. Sarbach,
Phys. Rev. Lett. {\bf 84}, 3033 (2000).

\bibitem{ershov}
A.A. Ershov and D.V. Gal'tsov,
Phys. Lett. {\bf {A150}}, 159 (1990).

\bibitem{breit}
P. Breitenlohner, P. Forgacs and D. Maison,
Comm. Math. Phys. {\bf {163}}, 141 (1994).

\bibitem{volkov}
M.S. Volkov, N. Straumann, G. Lavrelashvili,
M. Heusler and O. Brodbeck,
Phys. Rev. {\bf {D54}}, 7243 (1996).

\bibitem{kanti}
P. Kanti and E. Winstanley,
Phys. Rev. {\bf {D61}}, 084032 (2000).

\bibitem{SHB-PRD}
O. Sarbach, M. Heusler and O. Brodbeck,
{\em Self-adjoint wave equations for dynamical perturbations of self-gravitating fields},
gr-qc/0010067; to appear in Phys. Rev. {\bf D}.

\bibitem{AAL-CurvPert}
A. Anderson, A.M. Abrahams, and C. Lea,
Phys. Rev. {\bf D58}, 064015 (1998).

\bibitem{RW}
T. Regge and J. Wheeler,
Phys. Rev. {\bf 108}, 1063 (1957).

\bibitem{Volkov}
M.S. Volkov,
in {\it Proceedings of the Workshop ``Geometry and Integrable Models''},
edited by P.N. Pyatov and S.N. Solodukhin, pages 55-77
(World Scientific, 1996).

\bibitem{SHB-Odd}
O. Sarbach, M. Heusler, and O. Brodbeck,
Phys. Rev. {\bf D62}, 084001 (2000).

\bibitem{VG-trial}
M.S. Volkov and D.V. Gal'tsov,
Phys. Lett. B {\bf 341}, 279 (1995).

\bibitem{AQ-Nodal}
H. Amann and P. Quittner,
J. Math. Phys. {\bf 36}, 4553 (1995).

\bibitem{HP}
S.W. Hawking and D.N. Page,
Comm. Math. Phys. {\bf {87}}, 577 (1983).

\bibitem{zhou}
Z.-H. Zhou and N. Straumann,
Nucl. Phys. {\bf {B360}}, 180 (1991).

\bibitem{BHVS-Rotation}
M.S. Volkov and N. Straumann,
Phys. Rev. Lett. {\bf 79}, 1428 (1997);
\newline
O. Brodbeck, M. Heusler, N. Straumann, and M.S. Volkov,
Phys. Rev. Lett. {\bf 79}, 4310 (1997).



\end{thebibliography}
\end{document}